\magnification=\magstep1

\font\sm=cmss8 at 8truept
\font\smc=cmss8 scaled\magstep3

\font\tty=cmtt10 at 11truept

\def\Var{\hbox{\rm Var}}
\def\Cov{\hbox{\rm Cov}}

\tolerance=1000
\vsize=8.5truein
\hsize=5.5truein
\baselineskip=20truept
\parindent=6mm
\raggedright
\raggedbottom

\nopagenumbers
\pageno=1
\centerline{\bf Parsimony, Model Adequacy and Periodic Correlation}
\vskip 0.1truein
\centerline{\bf in Time Series Forecasting  }
\vskip 1.5truein
\centerline{A. Ian McLeod}
\centerline{Department of Statistical and Actuarial Sciences}
\centerline{The University of Western Ontario}
\centerline{London, Ontario N6A 5B7}
\centerline{Canada}
\vskip 1.5in

\noindent
McLeod, A.I., (1993), Parsimony, model adequacy and periodic correlation in forecasting time series, {\it International Statistical Review} 61/3, 387-393.
\vskip 0.5in
\noindent
{\bf Stable URL:} http://www.jstor.org.proxy1.lib.uwo.ca/stable/1403750 
\vskip 0.14in
{\bf DOI:} 10.2307/1403750 

\vfill\break\eject
\headline={\ifodd\pageno\rightheadline \else\leftheadline\fi}
\def\rightheadline{\sm\hfil A.I. McLeod\hfil\folio}
\def\leftheadline{\sm\folio\hfil Parsimony and Model Adequacy in Forecasting\hfil}

\line{\bf Summary\hfil}
\vskip .18truein
{The merits of the modelling philosophy of Box \& Jenkins (1970) are
illustrated with a summary of our recent work on seasonal river flow
forecasting.
Specifically, this work demonstrates that the principle of parsimony,
which has been questioned by several authors recently, is
helpful in selecting the best model for forecasting seasonal
river flow.
Our work also demonstrates the importance of model adequacy.
An adequate model for seasonal river flow must incorporate seasonal
periodic correlation.
The usual autoregressive-moving average (ARMA)
and seasonal ARMA models are not adequate in this
respect for seasonal river flow time series.
A new diagnostic check,
for detecting periodic correlation in fitted ARMA models is developed
in this paper.
This diagnostic check is recommended for routine use
when fitting seasonal ARMA models.
It is shown that this diagnostic check indicates
that many seasonal economic time
series also exhibit periodic correlation. Since the standard forecasting
methods are inadequate on this account, it can be concluded that
in many cases, the forecasts produced are sub-optimal.
Finally, a limitation of the arbitrary combination of forecasts
is also illustrated. Combining forecasts from an adequate parsimonious
model with an inadequate model did not improve the forecasts whereas
combining the two forecasts of two inadequate models did yield
an improvement in forecasting performance.
These findings also support the model building philosophy of
Box \& Jenkins.
The non-intuitive findings of
Newbold \& Granger (1974) and Winkler \& Makridakis (1983)
that the apparent arbitrary combination of forecasts from similar
models will lead to forecasting performance is not supported by our
case study with river flow forecasting.}


\vfill\break\eject
\line{\bf 1\quad Introduction \hfil}
\vskip .18truein
The main purpose of this paper is to
discuss some general statistical
principles
which are elucidated by our recent work in
river flow forecasting
(McLeod et al., 1987; Noakes et al., 1985; Thompstone
et al., 1985a).
Also based on these case studies, a new diagnostic check for periodic
correlation in the residuals of fitted ARMA models is developed.
This diagnostic check is suitable for routine use when fitting
seasonal ARMA models.

Briefly our experience with river flow time series
suggests that the best forecasting results are obtained by
following the general model building philosophy
implicit in Box \& Jenkins (1970) with suitable modifications and
improvements.
In general terms, this approach is iterative and advocates choosing
the most parsimonious adequate statistical model.
Two basic principles of special relevance are:
\break
\smallskip
\hbox{\smc Principle 1: Model Adequacy.\/} The model is considered
adequate if it incorporates all relevant information and if when
calibrated to the data, no important significant departures from
the statistical assumptions made can be found.
\break
\smallskip
\hbox{\smc Principle 2: Model Parsimony.\/}
The principle of parsimony means that
the simplest possible model should be chosen.
\smallskip
\smallskip

One can view the problem of statistical modelling as choosing
an adequate statistical model which is the most parsimonious.
In mathematical programming terminology we could say that
the problem of statistical modelling has an objective function which is
to minimize the model complexity (Model Parsimony)
subject to the constraint of
Model Adequacy.

In \S 2 the results of a case study of forecasting monthly
river flow time series is summarized. Here the importance
of incorporating periodic correlation in the forecasting
model is demonstrated.
For a seasonal time series denoted by $z_{r,p}$ where $r$ denotes
the year and $p$ denotes the seasonal period, the periodic correlation
coefficient is defined by
$$\rho_{m}(\ell) =
{\gamma_{m}(\ell)\over\surd (\gamma_m(0) \gamma_{m-\ell}(0))},$$
where
$$\gamma_{m}(\ell) = \Cov(z_{r,m}, z_{r,m-\ell}).$$

The concept of periodically correlated processes was introduced
by Gladyshev (1961).
The first application of periodic time series models
seems to have been by hydrologists Thomas \& Fiering (1962).
Since that time there have been very extensive developments in the theory and
applications of periodically correlated time series. For a review of
the probabilistic literature on periodically correlated processes,
see Yaglom (1986, \S 26.5; 1987).
Miamee (1990) and
Sakai (1991)
have derived new theoretical results and conditions on the spectral density
function of periodically correlated time series.
On the statistical methodology side, contributions
to periodically correlated time series modelling have been made by
Jones \& Brelsford (1967),
Moss \& Bryson (1974),
Pagano (1978),
Cleveland \& Tiao (1979),
Troutman (1979),
Tiao \& Grupe (1980),
Sakai (1982),
Dunsmuir, W. (1983),
Thompstone et al. (1985b),
Vecchia (1985a, 1985b),
Vecchia et al. (1985),
Li \&	Hui (1988),
Jim\'enez et al. (1989),
Hurd \& Gerr (1991),
Osborn, D.R. (1991)
and
Vecchia, A.V. \& Ballerini, R. (1991).
Periodic time series models are often used for modelling seasonal
time series -- especially environmetric series.
However, several
other interesting applications include multiple spectral estimation
(Newton, 1982) and multichannel signal processing (Sakai, 1990).

In some situations, as in the case study in \S 3,
a comprehensive modelling approach which satisfies both
adequacy and parsimony principles
may not be practical either for
reasons of expediency or because a suitable model cannot be found with
available methodology. In this case, we have found combined forecasts to
be useful. On the other hand, if a good model can be found, our experience
suggests that the forecast cannot be significantly improved by combining
it with forecasts from models which are less parsimonious or less
adequate. This latter result is at variance with the results reported by
Winkler \& Makridakis (1983) and Newbold \& Granger (1974). Perhaps
this is due to the fact that the river flow time series used in our studies
are generally longer and more homogeneous than the economic series used
by the aforementioned authors. The skill of the modeller in developing
an adequate model could also be a factor.

In order to make the hydrological data sets used in the case studies
referred to
in this paper readily accessible to other researchers, all data
is available in the StatLib archive {\tty riverflow}.

\vfill\break\eject
\vskip .3truein
\line{\bf 2\quad Monthly River Flow Case Study\hfil}
\medskip
The data in this case study (Noakes et al., 1985) consisted of thirty mean
monthly river flows for periods of from 37 to 64 years. Various models and model
selection procedures were used to calibrate a model to each data set
omitting the last three years of data. The one-step ahead forecasts were
then compared for the last three years (36 values). The best forecasts as
judged by the root mean-square error and other criteria were obtained with
the family of periodic autoregressive models.

The periodic autoregression model equation may be written
$$\phi_m(B)(Z_{r,m}-\mu_m)=a_{r,m}\eqno(1)$$
where $Z_{r,m}$ denotes the logarithmic flow for the $r^{th}$ year and
$m^{th}$ month, $\mu_m$ denotes the corresponding monthly mean,
$a_{r,m}, r=1,2,\ldots , m=1,2,\ldots, 12$
are a sequence independent normal random numbers
with mean zero and variance, $\sigma^2_m,$
and
$$\phi_m(B)=1-\phi_{m,1}B-\ldots -\phi_{m,p_m}B^{p_m}\eqno(2)$$
where $B$ is the backshift operator on $t$, where $t=12(r-1)+m$. Several model
selection techniques were used to select $p_m$ $(m=1,2,\ldots ,12)$. It
was found that a periodic autoregression
which was determined by choosing $p_m$
as small as possible to achieve an adequate fit gave the best forecasts.
This was accomplished by initially determining $p_m$ based on a plot of
the periodic partial autocorrelation function and then checking the
adequacy of the fitted model. Our approach is thus a natural extension
of that of Box \& Jenkins (1970).

On the other hand, a subset periodic autoregression approach was found to produce
comparatively very poor forecasts. In this approach, for each period
all possible autoregressions with some parameters constrained to
zero and with $p_m=12$ were examined ($2^{12}$ possibilities) and the best
model was selected with the Akaike Information Criterion
(Akaike, 1974) as well as the Bayes Information Criterion
(Schwarz, 1978).
It was also noticed that the
resulting models were always less parsimonious than that selected by
the first approach.

The seasonal ARMA model developed by Box \& Jenkins (1970, Ch. 9)
did not perform
very well either. In this case, the diagnostic check, developed in the
next \S 4, indicates that this is due to model inadequacy.

The periodic autoregression
and seasonal ARMA represent quite different families of time
series models. Not only do the models differ in the correlation structure
but the specification of seasonality is purely stochastic in the seasonal ARMA
model and purely deterministic in the case of the
periodic autoregression. Moreover neither
specification is likely to be absolutely correct.
Thus although the periodic autoregression model forecasted best
and was considered
to represent a more valid statistical model, it might be thought from
the experience reported by Newbold \& Granger (1974) and Winkler
\& Makridakis (1983) that combining the
periodic autoregression and seasonal ARMA forecasts
would be helpful. As shown in McLeod et al. (1987) this is not
the case. In particular with method 1 of Winkler \& Makridakis
(1983, p. 152) the periodic autoregression forecast had a smaller
mean square error
at least 17 times out of 30.
Thus combined forecasts cannot be
recommended in this situation.

\vfill\break\eject
\vskip .3truein
\line{\bf 3\quad Quarter-Monthly River Flow Case Study\hfil}
\vskip .18truein
The object of this study (Thompstone et al., 1985a)  was to obtain
one-step-ahead forecasts of the quarter-monthly, i.e. almost weekly,
inflows to the Lac St. Jean reservoir system operated by Alcan Limited.
Complete time series on past quarter-month inflows, precipitation and
snowmelt in the river basin were available for 30 years. A Box-Jenkins
multiple transfer-function noise model with precipitation and
snowmelt as inputs was found to provide an adequate fit to the
deseasonalized data in many respects except that it did not account
for the periodic correlation effect. A periodic autoregression
model was also fit but this
model did not take into account the covariates precipitation and snowmelt.
It could be suggested that at this stage a periodic-transfer-function
noise model should be developed to take into account both factors.
However such a model could easily involve too many parameters and,
in any case, it was not possible to calibrate it with our existing
computer software.
Perhaps future work will result in a suitable model.
Finally, a third model which was
a semi-theoretical hydrological model which incorporates various
hydrological and meteorological
information in a conceptual model of river flow. The conceptual modelling
approach has been strongly advocated by certain hydrologists who feel that time
series methods are too empirical.

All three models were calibrated on data for 27 years and then used to
produce one-step-ahead forecasts over the next three years (144 periods).
The root mean square error for transfer-function noise,
periodic autoregression and conceptual model for forecasting logarithmic
flows were respectively 0.2790, 0.3009 and 0.3894. When the forecasts
were combined by simple averaging the root
mean square error dropped to 0.1355.
More sophisticated combination techniques were found to lead to
even further improvements.

It is interesting to note that the empirical time series approach
outperformed the more theoretical conceptual approach which has
been strongly advocated by some hydrologists. A similar
phenomenon with macro-economic time series forecasting as previously
been found (Naylor et al., 1972).

\vfill\break\eject
\vskip .3truein
\line{\bf 4\quad A New Diagnostic Check For Periodic Autocorrelation\hfil}
\medskip
The seasonal ARMA model of order $(p,d,q)(p_s,d_s,q_s)_s$ may be written
$$\Phi (B^s)\phi (B)\nabla_s^{d_s}\nabla^dZ_t=\Theta (B^s)\theta
(B)a_t,\eqno(3)$$
where $Z_t$ is the observation at time $t$ and $a_t$
is a sequence of independent normal random variables with mean zero
and variance $\sigma^2$.
For monthly time series $s=12$ and $t=12(r-1)+m$, where
$r$ and $m$ represent the year and month respectively. The polynomials
$\Phi (B^s)$, $\phi (B)$, $\Theta (B^s)$ and $\theta (B)$ of degrees
$p_s$, $p$, $q_s$ and $q$ specify the autoregressive and moving
average components of the model. The terms $\nabla_s=1-B^s$
and $\nabla =1-B$ represent the seasonal and non-seasonal differencing
operators. Using standard model selection techniques (Box \& Jenkins, 1970;
Hipel et al., 1977) it was found that most monthly river flow
time series could be tentatively modelled as a seasonal ARMA model of order
$(p,0,1)(0,1,1)_{12}$, where $p=0$, $1$ or $2$.
The diagnostic check described below can be used to check for
model inadequacy due to periodic correlation in the residuals of
such fitted models.

The residual periodic autocorrelation at lag $k\ge 1$ may be written
$${\hat r}_m(k)={{\sum_r{\hat a}_{r,m}{\hat a}_{r,m-k}}\over
{\sqrt{\sum_r{\hat a}_{r,m}^2\sum_r{\hat a}^2_{r,m-k}}}},\eqno(4)$$
where ${\hat a}_{r,m}$ denotes the seasonal ARMA model residual for period
$t=12(r-1)+m$ $(r=1,\ldots ,N; m=1,\ldots ,12)$. If the seasonal ARMA
model is adequate then using the methodology in McLeod (1978)
it can be shown for any fixed $M\ge 1$, $\surd N{\bf{\hat r}}^{(m)}=
\surd N({\hat r}_m(1),\ldots ,{\hat r}_m(M))$ is asymptotically normal
with mean zero and covariance matrix $(1_M-Q/12)/N$, where
$1_M$ is the identity matrix of order $M$ and $Q=XI^{-1}X^T$, where
$X$ and $I$ are given in eq. (44) of McLeod (1978). Moreover,
${\sqrt N}{\bf r}^{(m)}$ and $\surd N{\bf r}^{(m^\prime )}$ are
asymptotically independent when $m\not= m^\prime$. Since the
diagonal elements of $Q$ are all less than one, it follows that
to a good approximation,
${\hat r}_m(1), m=1,\ldots ,12$ are jointly normally distributed
with mean vector zero, diagonal covariance matrix and
$\Var({\hat r}_m(1)) = N^{-1}.$
A diagnostic
check for detecting periodic autocorrelation in seasonal ARMA model
residuals is given by
$$S=N\sum^{12}_{m=1}{\hat r}^2_m(1)\eqno(5)$$
which should be approximately $\chi^2$-distributed on 12 df.

As a check on the asymptotic approximation involved, a brief simulation
experiment was performed. A $(1,0,0)(0,0,0)_{12}$ model with
$\phi_1=-0.9,-0.6,-0.3,0.3,0.6$ and $0.9$ was simulated. Table 1 summarizes
the results on $S$ for one thousand simulations with $N=17$. The
empirical significance level of a nominal 5\% test was estimated by
counting the number of times that $S$ exceeded 21.0261. From Table 1,
the approximation is seen to be adequate for practical purposes. In
further experiments with $N=34$ and 68, the approximation was seen to
improve although the empirical significance level was still
slightly less than 0.05 in all cases. This suggests that in general the
significance will be slightly overestimated. For example, if the
observed value of $S$ indicates significant periodic correlation at the
5\% level, the true significance level will be slightly less than 5\%.
\bigskip
\centerline{[Table 1 here]}
\bigskip
The data on the Saugeen River (1915--1976) is illustrative of the
usefulness of this new diagnostic check.
A $(1,0,1)$ $(0,$$1,$$1)_{12}$ model was fit to the logarithmic flows and
passed all diagnostic checks given in Box \& Jenkins (1970).
However, it was found that
$S=59.6$ indicating very significant residual periodic correlation.
As indicated in the next section, it appears that many
seasonal economic time series also exhibit such periodic residual
correlations.
\vfill\break\eject
\bigskip
\line{\bf Table 1\hfil}
\smallskip
\line{\it Empirical mean, variance and significance level of $S$\hfil }
\line{\it with $N=17$ in 1000 simulations using a nominal 5\% test.\hfil}

$$\vbox{\tabskip=1em plus2em minus.5em
\def\tablerule{\noalign{\hrule}}
\halign to\hsize{\hfil#&\hfil#&\hfil#&\hfil#\cr
\hidewidth $\phi_1$\hidewidth&\hidewidth{ Mean}\hidewidth
&\hidewidth{ Variance}\hidewidth&\hidewidth{ Significance}
\hidewidth\cr
&&\hidewidth&\hidewidth{ level}\hidewidth\cr
&&&\cr
\tablerule\cr
&&&\cr
--0.9&11.9&19.3&0.032\cr
--0.6&11.5&18.6&0.025\cr
--0.3&11.3&19.0&0.023\cr
0.0&10.9&16.6&0.016\cr
0.3&11.1&19.7&0.027\cr
0.6&11.5&18.6&0.026\cr
0.9&11.7&19.7&0.030\cr
&&&\cr
\tablerule\cr}}$$
\bigskip
\vfill\break\eject
\vskip .3truein

\line{\bf 5\quad Application to Forecasting Economic Time Series\hfil}
\medskip
Many seasonal economic time series may exhibit periodic correlation
which most of the standard approaches do not take into account.
The diagnostic check of \S 4 may be applied routinely when fitting
seasonal ARMA models. Table 2 shows the results of testing the
seasonal ARMA models
fitted by Miller \& Wichern (1977, p.432) to four Wisconsin series.
It is seen that in two out of the four series
there is very significant periodic
correlation.
In these cases, models which take this correlation
into account may be expected to produce improved forecasts.
A comprehensive new approach to the modelling and forecasting
of such series is
given by McLeod (1992).
\bigskip
\centerline{[Table 2 here]}
\bigskip
\vskip .3truein

\line{\bf Acknowledgements\hfil}
\medskip
This research was supported by NSERC.

\vfill\break\eject
\bigskip
\line{\bf Table 2\hfil}
\smallskip
\line{\it Diagnostic Test For Residual Periodic Correlation\hfil}
\smallskip
\line{\it For Four Wisconsin Series From Miller \& Wichern\hfil}
$$\vbox{\tabskip=1em plus2em minus.5em
\def\tablerule{\noalign{\hrule}}
\halign to\hsize{\hfil#&\hfil#&\hfil#&\hfil#\cr
\hidewidth Category\hidewidth&\hidewidth{\it S}\hidewidth
&\hidewidth{\it d.f.}\hidewidth&\hidewidth{ Significance}
\hidewidth\cr
&&\hidewidth&\hidewidth{ level}\hidewidth\cr
&&&\cr
\tablerule\cr
&&&\cr
Food Products\hfill&25.36&12&0.013\ \cr
Fabricated Metals\hfill&36.8&12&0.0002\cr
Transportation Equipment\hfill&11.6&12&0.478\ \cr
Trade\hfill&6.98&12&0.859\ \cr
&&&\cr
\tablerule\cr}}$$

\vfill\break\eject
\baselineskip=24truept
\line{\bf References\hfil}
\vskip .18truein
\def\ref{\noindent\hangindent 15pt}

\ref Akaike, H. (1974).
A new look at the statistical model identification.
{\it IEEE Trans. Autom. Control\/} {\bf 19\/}, 716--723.

\ref Box, G.~E.~P. \& Jenkins, G.~M. (1970). {\it Time Series
Analysis Forecasting and Control.} San Francisco: Holden-Day.

\ref Cleveland, W.P. \& Tiao, G.C. (1979).
Modeling seasonal time series.
{\it \'Economie Appliqu\'ee\/} {\bf 32,} 107--129.

\ref Dunsmuir, W. (1984).
Time series regression with periodically correlated errors and
missing data.
In {\it Time Series Analysis of Irregularly Observed Data\/},
Ed. E. Parzen. Springer-Verlag: New York.

\ref Hipel, K.~W., McLeod, A.~I. \& Lennox, W.~C. (1977). Advances
in Box-Jenkins modelling. {\it Water Resources Res.} {\bf 13,\/} 567--586.

\ref Hurd, H.L. and Gerr, N.L. (1991).
Graphical methods for determining the presence of periodic correlation.
{\it J. Time Ser. Anal.\/} {\bf 12,} 337--350.

\ref Glady\v sev, E.G. (1961).
Periodically correlated random sequences.
{\it Soviet Math. Dokl.\/} {\bf 2,} 385--388.

\ref Jim\'enez, C., McLeod, A.I. \& Hipel, K.W. (1989).
Kalman filter estimation for periodic autoregressive-moving average
models.
{\it Stochastic Hydrology and Hydraulics\/} {\bf 3,} 227-240.

\ref Jones, R.H. \& Brelsford, W. (1967).
Time series with periodic structure.
{\it Biometrika\/} {\bf 54,} 403--408.

\ref Li, W.K. \& Hui, Y.V. (1988).
An algorithm for the exact likelihood of periodic autoregressive
moving average models.
{\it Commun. Statist. Simulation Comput.\/} {\bf 14\/}, 1483--1494.

\ref Miamee, A.G. (1990).
Periodically correlated processes and their stationary dilations.
{\it SIAM, J. Appl. Math.\/} {\bf 50,} 1194--1199.

\ref McLeod, A.~I. (1978). On the distribution of residual
autocorrelations in Box-Jenkins models.
{\it J.~R. Statist. Soc.\/} B {\bf 40,} 296--302.

\ref McLeod, A.~I., Noakes, D.~J., Hipel, K.~W. \& Thompstone, R.~M.
(1987). Combining hydrological forecasts. {\it J. Water
Resourc. Planning \& Manage. Div. Proc. ASCE\/} {\bf 113,} 29--41.

\ref McLeod, A.~I. (1992, to appear).
An extension of Box-Jenkins seasonal models.

\ref Miller, R.~B. \& Wichern, D.~W. (1977).
{\it Intermediate Business Statistics.\/}
New York: Holt, Reinhart and Winston.

\ref Moss, M.E. and Bryson, M.C. (1974).
Autocorrelation structure of monthly streamflows.
{\it Water Resources Res.\/} {\bf 10,} 733--744.

\ref Naylor, T. II, Seaks, T.G. \& Wichern, D.W. (1972).
Box-Jenkins methods: An alternative to econometric models.
{\it Int. Statist Rev.\/} {\bf 40,} 123--137.

\ref Newbold, P. \& Granger, C.~W.~J. (1974). Experience with
forecasting univariate time series and the combination of forecasts.
{\it J.~R. Statist. Soc.\/} A  {\bf 137,} 131--165.

\ref Newton, H.J. (1982).
Using periodic autoregressions for multiple spectral estimation.
{\it Technometrics\/} {\bf 24,} 109--116.

\ref Noakes, D.~J., McLeod, A.~I. \& Hipel, K.~W. (1985). Forecasting
monthly riverflow time series.
{\it Int. J. Forecast.} {\bf 1,} 179--190.

\ref Osborn, D.R. (1991).
The implications of periodically varying coefficients for seasonal
time series processes.
{\it J. Econometrics\/} {\bf 48,} 373--384.

\ref Pagano, M. (1978). On periodic and multiple autoregressions.
{\it Ann. Statist.\/} {\bf 6,\/} 1310--1317.

\ref Sakai, H. (1982).
Circular lattice filtering using Pagano's method.
{\it IEEE Trans. Acoust. Speech Signal Process.\/} {\bf 30,}
279--287.

\ref Sakai, H. (1990).
Cirular lattice filtering for recursive least squares and ARMA
modeling.
In {\it Linear Circuits, Systems and Signal Processing,\/}
Ed. N. Nagai, New York: Dekker.

\ref Sakai, H. (1991).
On the spectral density matrix of a periodic ARMA process.
{J. Time Ser. Anal.\/} {\bf 12,} 73--82.

\ref Schwarz, G. (1978).
Estimating the dimension of a model.
{\it Ann. Statist.\/} {\bf 6\/}, 461--464.

\ref Thomas, H.~A. \& Fiering, M.~B. (1962). Mathematical synthesis of
stream flow sequences for the analysis of river basins by simulation.
In {\it Design of Water Resources\/}, Ed. Maass, A., Hufschmidt, M.~M.,
Dorfman, R., Thomas, H.~A., Marglin, S.~A. \& Fair, G.~M.
Harvard University Press.

\ref Tiao, G.C. \& Grupe, M.R. (1980).
Hidden periodic autoregressive-moving average models in time series
data.
{\it Biometrika\/} {\bf 67,} 365--73.

\ref Thompstone, R.~M., Hipel, K.~W. \& McLeod, A.~I. (1985a). Forecasting
quarter-monthly riverflow. {\it Water Resources Bull.} {\bf 21,} 731--741.

\ref Thompstone, R.~M., Hipel, K.~W. \& McLeod, A.~I. (1985b).
Simulation of monthly hydrological time series.
In {\it Stochastic Hydrology\/}. Ed. A~.I. McLeod, Dordrecht: Reidel.

\ref Troutman, B.M. (1979).
Some results in periodic autoregression.
{\it Biomet\-rika\/} {\bf 66,} 219--228.

\ref Vecchia, A.V. (1985a).
Maximum likelihood estimation for periodic autoregressive-moving average
models.
{\it Technometrics\/} {\bf 27,} 375--384.

\ref Vecchia, A.V. (1985b).
Periodic autoregressive-moving average modeling with applications to
water resources.
{\it Water Resources Bull.\/} {\bf 21,} 721-730.

\ref Vecchia, A.V., Obeysekera, J.T., Salas, J.~D. \& Boes, D.~C. (1985).
Aggregation and estimation for low-order periodic ARMA models.
{\it Water Resources Res.\/} {\bf 19,} 1297--1306.

\ref Vecchia, A.V. \& Ballerini, R. (1991).
Testing for periodic autocorrelations in seasonal time series data.
{\it Biometrika\/} {\bf 78,} 53--63.

\ref Winkler, R.~L. \& Makridakis, S. (1983). The combination of forecasts.
{\it J.~R. Statist. Soc.} A {\bf 146,} 150--157.

\ref Yaglom, A.M. (1986).
{\it Correlation Theory of Stationary and Related Random Functions I.
Basic Results\/}.
New York: Springer-Verlag.

\ref Yaglom, A.M. (1987).
{\it Correlation Theory of Stationary and Related Random Functions II.
Supplementary Notes and References\/}.
New York: Springer-Verlag.

\vfill\break\eject
\baselineskip=20truept
\parindent=6mm
\line{\bf R\'esum\'e\/}
Les m\'erites de la philosophie de la mod\'elisation de Box et
Jenkins (1970) sont illustr\'ees par un r\'esum\'e de nos recherches
r\'ecentes sur la pr\'evision de saisonniers d\'ebits en rivi\`ere.
En particulier, nos r\'esultats d\'emontrent que le principe de la
parcimonie, que plusieurs auteurs ont mis en question, est utile
\`a la s\'election du meilleur mod\`ele pour pr\'evoir les saisonniers
d\'ebits en rivi\`ere. Nos recherches d\'emontrent l'importance de la
comp\'etence d'un mod\`ele. Un mod\`ele ad\'equat de saisonniers
d\'ebits en rivi\`ere doit incorporer la corr\'elation p\'eriodique
et saisonni\`ere. Les mod\`eles autor\'egressifs \`a moyenne mobile
(ARMA) habituels et les mod\`eles ARMA saisonniers ne sont pas ad\'equats
\`a cet \'egard pour les s\'eries de saisonniers d\'ebits en rivi\`ere.
Dans cet article, on d\'eveloppe une nouvelle m\'ethode pour d\'eceler la
corr\'elation p\'eriodique dans les mod\`eles ARMA ajust\'es.
Cette m\'ethode est \`a utiliser habituellement dans l'ajustement des
mod\`eles ARMA saisonniers.
Cette m\'ethode indique que beaucoup de s\'eries chronologiques
\'economiques font preuve de la corr\'elation p\'eriodique aussi.
Puisque les m\'ethodes ordinaires de pr\'evision ne sont pas ad\'equats,
on peut conclure que dans beaucoup de cas les pr\'evisions produites
sont moins qu'optimales.
En dernier lieu, une limitation \`a la combinaison arbitraire des
pr\'evisions est illustr\'ee aussi.
La combinaison des pr\'evisions d'un mod\`ele parcimonieux et ad\'equat
avec celles d'un mod\`ele inad\'equat n'am\'eliora pas les
pr\'evisions.
Cependant, le fait de combiner les deux pr\'evisions de
deux mod\`eles inad\'equats produisit une am\'elioration de la performance
de la pr\'evision.
Ces r\'esultats appuient aussi la philosophie de la mod\'elisation de Box
et Jenkins.
Les r\'esultats non-intuitifs de Newbold et Granger (1974) et de
Winkler et Makridakis (1983) indiquent que la combinaison apparente
et arbitraire de pr\'evisions de mod\`eles semblables men\`era \`a
la performance des pr\'evisions.
Cette conclusion n'est pas soutenue par notre \'etude de cas
portant sur la pr\'evision des d\'ebits en rivi\`ere.

\end